# Fine temporal structure of cardiorespiratory synchronization

Sungwoo Ahn[1], Jessica Solfest[1], and Leonid L Rubchinsky[1,2]

[1] Department of Mathematical Sciences and Center for Mathematical Biosciences, Indiana University Purdue University Indianapolis, Indianapolis, IN 46202

Phone (317) 274-6918  Fax (317) 274-3460

[2] Stark Neurosciences Research Institute, Indiana University School of Medicine, Indianapolis, IN 46202

Running head: cardiorespiratory synchronization

Corresponding author: Leonid L Rubchinsky

Department of Mathematical Sciences, Indiana University Purdue University Indianapolis

402 N. Blackford St., Indianapolis, IN 46202, USA.

Email: leo@math.iupui.edu

Phone (317) 274-6918

Fax (317) 274-3460

**Abstract**

Cardiac and respiratory rhythms are known to exhibit a modest degree of phase synchronization, which is affected by age, diseases, and other factors. We study the fine temporal structure of this synchrony in healthy young, healthy elderly, and elderly subjects with coronary artery disease. We employ novel time-series analysis to explore how phases of oscillations go in and out of the phase-locked state at each cycle of oscillations. For the first time we show that cardiorespiratory system is engaged in weakly synchronized dynamics with a very specific temporal pattern of synchrony: the oscillations go out of synchrony frequently, but return to the synchronous state very quickly (usually within just one cycle of oscillations). Properties of synchrony depended on the age and disease status. Healthy subjects exhibited more synchrony at the higher (1:4) frequency-locking ratio between respiratory and cardiac rhythms, while subjects with coronary artery disease exhibited relatively more 1:2 synchrony. However, multiple short desynchronization episodes prevailed regardless of the age and disease status. The same average synchrony level could alternatively be achieved with few long desynchronizations, but this was not observed in the data. This implies functional importance of short desynchronization dynamics. These dynamics suggest that a synchronous state is easy to create if needed, but is also easy to break. Short desynchronization dynamics may facilitate the mutual coordination of cardiac and respiratory rhythms by creating intermittent synchronous episodes. It may be an efficient background dynamics to promote adaptation of cardiorespiratory coordination to various external and internal factors.

**Introduction**

Synchronization between heartbeat and respiration has been extensively studied to understand function and dysfunction of cardiorespiratory interactions in health and disease (4, 10, 11, 14, 24, 26, 31, 33, 34, 35, 43, 44). Different linear and nonlinear data analysis methods and mathematical models have been used to characterize cardiorespiratory synchronization and underlying mechanisms (8, 9, 11, 14, 19, 24, 26, 35, 45). One of the manifestations of cardiorespiratory coupling is respiratory sinus arrhythmia (RSA), a well-known specific heartbeat variability within a respiratory cycle (e.g., 3, 33, 44). As was emphasized in (34), RSA is a modulation phenomenon (cardiac rhythm is modulated with respect to respiratory rhythm), as opposed to another form of rhythmic coordination: cardiorespiratory phase synchronization. While term synchronization is sometimes used in reference to RSA, modulation (such as RSA) and phase synchronization are different phenomena, and RSA and cardiorespiratory synchronization appear to be not correlated (4, 10, 34). Phase synchronization may be observed without rhythm modulation, like RSA. The synchrony between heartbeat and respiratory signals, although less visible in the data to the naked eye, has been previously studied at different frequency-locking ratios (4, 16, 24, 31, 34, 35, 43, 45). Phase synchrony and RSA may be generated by different physiological mechanisms (34) and have different functions. Phase synchronization in the cardiorespiratory system is less studied than RSA. The objective of the present study is to investigate the temporal properties of cardiorespiratory synchronization to facilitate further understanding of its physiological mechanisms and functions as one of the modes of cardiorespiratory interactions.

Previous approaches studied synchrony strength as an average measure computed for temporal intervals of some sufficiently long durations. However, the temporal patterns of synchrony in each cycle of oscillations (provided that some overall synchrony is present) have not been explored. This is the focus of the present study.

The interactions between the cardiovascular system and respiration involve a large number of feedback and feedforward mechanisms (e.g., 9, 21). In particular, baroreflex-mediated mechanisms may contribute to phase coordination between rhythms (e.g., 41). The phases of both rhythms can be locked with different ratios $n:m$ over time (4, 16, 24, 31, 34, 35). The frequency-locking ratio of $1:m$ appears to be the dominant ratio in certain physiological states and conditions including sleep states and resting state (4, 14, 25, 40, 43). Properties of cardiorespiratory synchronization (including the frequency ratio and the average synchronization strength between heartbeat and respiration) depend on age, gender, and physiological state (4, 16, 24, 26, 43). Aging may alter cardiac and regulatory dynamics, and reduce a person's ability to adapt to stress (15, 22). Syncope subjects show significant changes in cardiovagal baroreflex, arterial pressure, blood pressure, respiration, and R-R intervals (23, 26). The complexity of heart rate dynamics also depends on age and gender (17, 32). Baroreflex sensitivity and the influence of respiration on heart rate decrease with age (14, 20). However, certain temporal scaling, nonlinear properties of the cardiac system remain stable and intact, regardless of subjects' physiological conditions (6, 36). Thus, while aging and other physiological states substantially affect cardiorespiratory dynamics, there may be invariant temporal properties across all conditions that support the systems' adaptability.

In this study, we investigated temporal patterns of synchronization between heartbeat and respiration in healthy young, healthy elderly, and elderly subjects with coronary artery disease.

Synchrony between the cardiovascular and respiratory systems is usually weak or moderate (e.g., 24, 31, 34, 43), meaning that two coupled rhythms spend a substantial amount of time in a non-synchronous state. This justifies the study of synchronization/desynchronization transitions. If synchronous dynamics in the human cardiorespiratory system are necessary to facilitate its functionality, then there may be universal structure in cardiorespiratory synchrony for all subject groups and conditions. We employed novel time-series analysis techniques to explore how cardiac and respiratory rhythms in humans become transiently phase locked and unlocked over time. This approach was recently developed to characterize the fine temporal structure of synchronization with 1:1 frequency locking ratio (1). In the present study, we extended this method to the frequency locking ratio of $1:m$ where $m$ is the number of heartbeats within one respiratory cycle.

## Materials and Methods

*Subjects and data*. We used two different publicly available data sets: Massachusetts General Hospital/Marquette Foundation (MGH/MF) dataset (42) and Fantasia dataset (15). Both datasets are available at Physionet (12, 29). Subjects in the Fantasia dataset consisted of twenty young (21 - 34 years old) and twenty elderly (68 - 85 years old) healthy subjects. We took a subset of the MGH/MF dataset consisting of all elderly patients with coronary artery disease of the same age range as in the Fantasia database and for whom electrocardiogram (ECG) and respiratory impedance recordings were available. This totalled 74 elderly subjects (68 - 85 years old) in the MGH/MF group. In the following analysis, we considered three subject groups: MGH/MF elderly subjects with coronary artery disease (MO), Fantasia healthy elderly subjects (FO), and Fantasia healthy young subjects (FY).

ECG and respiratory impedance were collected at the sampling rate of 360 Hz for the MGH/MF dataset and at the sampling rate of 250 Hz for the Fantasia dataset. In our analysis, we used the times of R-peaks in the ECG, and changed ECG recording into binary signal (0 and 1).

*Phase-locking analysis*. The analysis of the phase domain is appropriate for the study of weakly or moderately synchronized dynamics (13, 30). In general, when coupling strength between oscillators increases from zero, synchrony may be observed first in the phase domain while the amplitudes of coupled oscillators may remain uncorrelated (30). That is, as the coupling strength between two oscillators increases from zero, the phases are entrained first, followed by amplitude correlations for stronger coupling. However, stronger coupling does not destroy phase correlations. Thus, phase analysis appears to be an efficient and sufficiently sensitive approach to study moderate levels of cardiorespiratory synchrony.

The peak frequencies of ECG and respiratory signals were computed from the power spectrum density (PSD) for a 7-min interval. Note that the PSDs of ECG signal based on the spectrum of counts method and spectrum of intervals method have similar frequency peaks (7). Moreover, our analysis is based on only the phases of signals (see below), which further eliminated qualitative difference between both spectrum methods. We set the cutoff ranges for the frequencies of ECG and respiratory signals as follows: ECG frequency is in [0.5, 2] Hz and respiratory frequency is in [0.15, 1] Hz. If either frequency is out of this range, then the time interval was excluded in the following analysis. This provides sufficiently limited spectral range so that the phase can be well defined, and assists in the removal of artifacts associated with very high and very low frequencies. The filtered frequency band was selected as $f_{peak} \pm f_{peak}/2$, where $f_{peak}$ is the peak frequency of each signal. The signals were filtered in the given frequency domain with Kaiser windowed digital FIR filter sampled at 360 Hz for the MGH/MF dataset and at 250 Hz for the Fantasia dataset. Zero-phase filtering was implemented to avoid phase distortions. The signal was represented as the Gabor representation via Hilbert transformation and projected onto the unit circle to extract the phase (13, 30), resulting in time-series of phases of respiratory and ECG signals, $\varphi_1(t)$ and $\varphi_2(t)$, for each subject. To clarify, let $x(t)$ be the filtered signal at the given frequency band. Then the complex analytic extension of $x(t)$ is given by

$$\zeta(t) = x(t) + i\,\bar{x}(t),$$

where $\bar{x}(t)$ is given by the Hilbert transform of the signal

$$\bar{x}(t) = H(x) = \frac{1}{\pi}\text{p.v.}\int_{-\infty}^{\infty}\frac{x(\tau)}{t-\tau}d\tau.$$

Then the analytic signal is projected on the unit circle

$$z(t) = \frac{\zeta(t)}{\|\zeta(t)\|} = e^{i\varphi(t)},$$

where $\|\zeta(t)\|$ is the modulus of $\zeta(t)$. Now one can extract the phase $\varphi(t)$ through the argument (angle) of $z(t)$. Note that it is convenient to define the phase as the accumulated angle over time so that one cycle is from 0 to $2\pi$, two cycles from $2\pi$ to $4\pi$, and so on. Fig. 1A shows an example of PSDs from the ECG (left panel) and respiratory impedance (right panel) from a subject in the FY group. Note the prominent peaks around 0.37 Hz for respiratory impedance and 1.13 Hz for ECG, resulting in the 1:3 frequency-locking ratio. When $1:m$ frequency-locking ratio held (allowing for a small deviation: $|f_{ECG}/f_{resp} - m| < 0.2$), we included the pair of signals from the subject in the analysis. In the following analysis, we considered the frequency-locking ratios of 1:2, 1:3, and 1:4. These ratios were seen in 64 MO subjects, 18 FO, and 17 FY subjects (not at the same time). Some subjects had different frequency-locking ratios at different time epochs. All time epochs were used to analyze the corresponding frequency-locking ratios.

For each pair of the time-series of phases $\varphi_1(t)$ and $\varphi_2(t)$, the condition for the $1:m$ phase-locking is $|\varphi_1(t_j) - m\varphi_2(t_j)| < const$, where $m$ is an integer and $const < 2\pi$. We utilized a widely used phase synchrony strength index:

$$\gamma = \left\| \frac{1}{M} \sum_{j=1}^{M} e^{i\theta(t_j)} \right\|^2,$$

where $\theta(t_j) = \varphi_1(t_j) - m\varphi_2(t_j)$ is the phase difference, $t_j$ is the time point for $j=1,..., M$, and $M$ is the number of sampling data points in the given time interval. This phase synchrony index may vary from 0 (no phase-locking) to 1 (perfect phase-locking).

*First-return map analysis of fine temporal structure of synchronization.* The phase synchronization index discussed above, as well as other similar measures, characterizes an average synchrony level while excluding temporal fluctuations of phase locking strength (1). The fine temporal structure of the synchronization was studied here with a recently developed approach based on the analysis of phase synchrony on short time scales via the first-return map (1). This approach was tested in various coupled oscillatory systems, including coupled Rössler oscillators, which exhibit intermittent synchrony, and was shown to reveal the temporal pattern of weak intermittent synchrony as it appears and disappears in time (1). We generalized this approach for $1:m$ phase locking in the following way. Whenever the slower (respiratory) phase $\varphi_1$ changed from negative to positive values, we recorded the other (faster, ECG) phase $\varphi_2$, generating a set of consecutive phase values $\{\phi_i\}$, $i = 1, ..., N$. Thus $\phi_i$ represents the phase difference between two signals at every $m$ cycles of the faster rhythm. A Kolmogorov-Smirnov test was performed to detect non-uniform distribution of $\{\phi_i\}_{i=1}^{N}$ at a 0.05 significance level to include signal pairs exhibiting synchronized episodes in the further analysis. The distribution of $\phi_i$ is usually unimodal and its peak is the preferred phase difference. Note that in a more general $n:m$ phase locking case (where both $n, m > 1$ are integers), this particular generalization of 1:1 method may not necessarily work.

Then $\phi_{i+1}$ vs. $\phi_i$ was plotted for $i = 1, \ldots, N-1$ (Fig. 1B). The phase space was partitioned into four regions with symmetric partitioning to allow for simple computation of the transitions between regions (Fig. 1B). We shifted all values of the phases so that the preferred phase difference was located in the center of the first region (quadrant). The first region is considered to be the synchronized state, as it corresponds to the preferred phase lag between oscillations, and all other regions are considered to be desynchronized states. We quantified transitions between different regions of the $(\phi_i, \phi_{i+1})$ space to describe how the system leaves and returns to the synchronous state.

[Figure 1]

The transition rates $r_{1,2,3,4}$ between four regions of the first-return map are defined as the number of points leaving a region, divided by the total number of points in that region. For example, $r_1$ is the number of points leaving the region I for the region II divided by the total number of points in the region I.

Fig. 1C shows examples of raw data and corresponding filtered data from ECG and respiratory signals (ECG signals are represented by bars of unit height corresponding to R-peaks). In the considered first-return map approach, the duration of desynchronization events is the number of steps that system spends away from the first region minus one. This is the number of cycles of oscillations the signals are not in the preferred phase relationship. For example, a desynchronization event lasting for one cycle of oscillations (Cycle 1) corresponds to the shortest path II→IV→I (see Fig. 1D). Two cycles of desynchronization events (Cycle 2) corresponds to the path II→III→IV→I and three cycles of desynchronization events (Cycle 3) corresponds to either II→III→III→IV→I or II→IV→II→IV→I (see Fig. 1D). Longer lengths of desynchronization events have more path options. The gray box in the lower panel of Fig. 1C shows an example of a desynchronization event with a length of one cycle (in sync - out of sync - in sync). To explore the properties of the fine temporal structure of the dynamics, we computed the relative frequencies (probabilities) of desynchronization events of different durations. Here, $Prob(Cycle\ n)$ is the probability of desynchronization events lasting for $n$ cycles of oscillations. We also considered the mean length of desynchronization events for each pair of signals

$$<l> = \frac{1}{L}\sum_{i=1}^{L} n_i ,$$

where $n_i$ is the duration of each desynchronization event and $L$ is the total number of desynchronization events.

Data analysis was performed in MATLAB (Mathworks, Nautick, MA) and R (www.r-project.org). Unless specified otherwise, all comparisons were first subjected to one-way analysis of variance testing (ANOVA) for groups as a factor. Further analysis was performed with t-test, Tukey's post-hoc test, and two proportion z-test at the significance level of *α = 0.05* or *α = 0.0125* (see the text below).

**Results**

We considered the distributions of desynchronization event durations, transition rates $r_{1,2,3,4}$, synchrony index γ, and mean length of desynchronization events $<l>$ for MGH/MF elderly subjects (MO), Fantasia elderly subjects (FO), and Fantasia young subjects (FY) at the frequency-locking ratios of 1:2, 1:3, and 1:4. Note that while some subjects exhibited 1:5 and 1:6 frequency-locking ratios, these dynamics were not in phase-synchrony, and the analysis of synchronization properties did not apply to these cases. A 1:1 frequency-locking ratio sometimes occurred in MO, but because this did not occur in healthy subjects, we also excluded these episodes from analysis. Nevertheless, while 1:1 episodes were rare, short desynchronization events were prevalent, similar to other frequency ratios described below.

*Short desynchronization events prevail in cardiorespiratory synchronization*

Analysis of desynchronization epochs in the first-return map revealed that desynchronization episodes tend to quickly return to the synchronous state of the preferred phase difference. For each pair of ECG and respiratory signals, desynchronization event durations were not uniformly distributed [$\chi^2$ goodness of fit test, *p<2.27e-16*]. We further determined which length of cycles was significantly different from others. For each frequency-locking ratio, we observed that the probability of desynchronizations lasting for one cycle of oscillations ("Cycle 1") was higher than those of longer durations, including the cumulative probability of all durations longer than 6 cycles ("Cycle>6"). This was statistically confirmed at the 1:2 frequency-locking ratio [one-sided paired t-test, *p<8.95e-53* for MO; Fig. 2A] and at the 1:3 and 1:4 ratios [one-sided paired t-test, *p<8.67e-9* for each subject group MO, FO, and FY; Figs. 3A and 4A]. Note that at the 1:2 frequency-locking ratio, there was not enough statistical power to distinguish the difference between probabilities of FO and FY due to the small sample sizes.

      Overall, the probability that the mode of the distribution of durations was at Cycle 1 was *85.14%*. Thus, regardless of subject groups and frequency-locking ratios, shortest desynchronization events (one cycle of oscillations) were dominant in cardiorespiratory synchronization. This implies that while cardiac and respiratory rhythms may go out of a phase-locked state frequently during overall synchronized dynamics, they usually do so for short time

intervals. As it will be clear from the details reported in the next subsection, this held for each considered frequency-locking ratio and group of subjects.

*Different frequency-locking ratios and subject groups exhibit different temporal dynamics*

We considered the fine temporal structure of synchronization/desynchronization for each group of subjects at different frequency-locking ratios. As we mentioned above, the mode of the distributions of desynchronization event durations was Cycle 1 with a very high probability (*85.14%*). Thus we considered the probability of Cycle 1 ($Prob(Cycle\ 1)$) as the representative measure among all durations of desynchronization events. This was complemented by the mean duration of desynchronization events $<l>$. Conversely, rate $r_1$ is related to the mean duration of synchronized episodes and thus characterizes synchronization, rather than desynchronization (1). The mean duration of synchronized episodes equals $1/r_1$ if the duration is measured as the number of iterations of the map $\phi_i$, which is essentially the number of oscillatory cycles (1). Average synchrony strength, γ, incorporates both synchronized and desynchronized episodes and thus characterizes average synchronous dynamics. In the following analysis we used four measures to study differences among the three groups: $Prob(Cycle\ 1), r_1, γ,$ and $<l>$. Note that, despite their overall similarity, these measures significantly varied across different groups and different frequency-locking ratios. This is expected because cardiac and respiratory rhythms synchrony in these cases may reflect different underlying physiology. To compare the distributions of these measures among the three groups, we first employed a one-way analysis of variance (ANOVA) test for groups as a factor for each measure at each frequency-locking ratio considered (1:2, 1:3, and 1:4).

We first considered the effect of groups at the 1:2 frequency-locking ratio. This ratio episodes included 26 subjects in MO, 3 subjects in FO, and 3 subjects in FY. Significant main effects of groups were observed for $Prob(Cycle\ 1)$ [$F(2,178)=4.35, p=1.44e-2$; Fig. 2A], $r_1$ [$F(2,178)=4.34, p=1.45e-2$; Fig. 2B], $\gamma$ [$F(2,178)=3.44, p=3.42e-2$; Fig. 2C], and $<l>$ [$F(2,178)=3.34, p=3.76e-2$; Fig. 2D]. We performed post-hoc tests to find the differences between groups. Significant differences between MO and FO were observed for $Prob(Cycle\ 1)$ [Tukey HSD, $p=1.36e-2$; Fig. 2A] and $<l>$ [Tukey HSD, $p=2.95e-2$; Fig. 2D]. Moreover, the significant difference between MO and FY was observed for $r_1$ [Tukey HSD, $p=3.13e-2$; Fig. 2B]. No differences of $\gamma$ among the three groups were observed [Tukey HSD, $p>0.05$; Fig. 2C]. Overall, this suggests that while average synchrony levels did not differ statistically, desynchronization episodes in MO tended to be shorter (and thus more numerous) than in healthy subjects. Note that few subjects in FO and FY were included at the 1:2 frequency-locking ratio because this ratio was rarely observed in the healthy subjects and the statistics did not have sufficient power to discriminate the difference between FO and FY.

[Figure 2]

The 1:3 frequency-locking ratio episodes included 43 subjects in MO, 14 subjects in FO, and 9 subjects in FY. Significant main effects of groups were observed for $Prob(Cycle\ 1)$ [$F(2,616)=13.27, p=2.29e-6$; Fig. 3A], $r_1$ [$F(2,616)=66.54, p<2.0e-16$; Fig. 3B], $\gamma$ [$F(2,616)=106.95, p<2.0e-16$; Fig. 3C], and $<l>$ [$F(2,616)=16.78, p=8.01e-8$; Fig. 3D]. We performed post-hoc tests to find the differences between groups. $Prob(Cycle\ 1)$ of FO was significantly different from those of MO and FY [Tukey HSD, $p\leq3.17e-2$; Fig. 3A], and $r_1$ of MO significantly differed from those of FO and FY [Tukey HSD, $p\leq2.95e-10$; Fig. 3B]. Significantly different $\gamma$ among the three groups were observed [Tukey HSD, $p\leq8.94e-6$; Fig.

3C] and $<l>$ of FY significantly differed from those of MO and FO [Tukey HSD, $p \leq 1.38e\text{-}4$; Fig. 3D]. At this frequency-locking ratio, MO had a significantly higher $r_1$ and lower γ than those of FO and FY. Thus, the MO group (old diseased subjects) displayed less synchronous dynamics than the FY group (healthy young subjects) at the 1:3 ratio. Furthermore, FY had a significantly higher γ, higher $Prob(Cycle\ 1)$, and lower $<l>$ than those of FO, indicating more synchronous dynamics in healthy young subjects at this frequency-locking ratio. Nonetheless, short desynchronizations were prevalent in all groups.

[Figure 3]

The 1:4 frequency-locking ratio episodes included 14 subjects in MO, 12 subjects in FO, and 12 subjects in FY. Significant main effects of groups were observed for $r_1$ [$F(2,247)=5.37$, $p=5.20e\text{-}3$; Fig. 4B] and γ [$F(2,247)=11.79$, $p=1.29e\text{-}5$; Fig. 4C], but not for the other two measures [$F(2,247) \leq 1.57$, $p>0.05$; Figs. 4A and 4D]. We performed post-hoc tests to find the differences between groups. We observed significant difference for $r_1$ between FO and FY [Tukey HSD, $p=3.56e\text{-}3$; Fig. 4B]. The synchrony index γ of FY was also significantly different from those of MO and FO [Tukey HSD, $p \leq 1.09e\text{-}3$; Fig. 4C]. At this frequency-locking ratio, FY had a significantly higher γ than those of MO and FO, suggesting more synchronous dynamics in healthy young subjects at this ratio. The desynchronization properties were similar among all three groups.

[Figure 4]

We then compared the average durations of synchronous and desynchronous states within groups at each frequency-locking ratio. That is, we compared $1/r_1$ and $<l>$ within each group at 1:2, 1:3, and 1:4 ratios. The durations of desynchronous states ($<l>$) were significantly longer than the durations of synchronous states ($1/r_1$) for all cases except MO at 1:2 and FY at

1:3 (paired t-test, $p\leq7.06e-4$). For these two exceptions, there was no difference between $1/r_1$ and $<l>$ (paired t-test, $p>0.05$). These results are reasonable because the overall levels of synchrony were high, therefore the durations of synchronous/desynchronous states were balanced. At higher frequency-locking ratios (such as 1:5 or higher), we also expect $<l>$ to be longer than $1/r_1$. However, higher ratios showed substantially less overall synchrony than the lower frequency-locking ratios, making our data analysis approach irrelevant.

The values of all four measures ($Prob(Cycle\ 1), r_1, \gamma, <l>$) varied depending on the frequency-locking ratios and subject groups, but the average levels of synchrony for all cases were weak or moderate. The shortest desynchronization events (one cycle of oscillations) were dominant everywhere. At the 1:2 and 1:3 frequency-locking ratios, MO significantly differed from both FO and FY. At the 1:3 and 1:4 frequency-locking ratios, FY was significantly different from FO. Lastly, at the 1:4 frequency-locking ratio, FY was significantly different from MO and FO.

Finally, we related the above four measures of synchrony to heartbeat and respiratory rates. To simplify the analysis we considered all four synchronization measures within all groups. Here, we normalized $<l>$ by dividing it by its maximum value. This makes $<l>$ vary from 0 to 1, like the other measures. The linear regression analysis was performed for the dependence of each synchronization measure on heartbeat and respiratory rates. The four measures of synchronization displayed no correlation or very weak correlation to heartbeat and respiratory rates ($R^2 \leq 2.18e-1$, $|slope|\leq5.31e-1$ for MO; $R^2 \leq 7.45e-2$, $|slope|\leq2.93e-1$ for FO; $R^2 \leq 8.18e-2$, $|slope|\leq1.58e-1$ for FY). These small $R^2$s and small slope values suggest that cardiac and respiratory rates are nearly independent of dynamical measures ($Prob(Cycle\ 1), r_1, \gamma, <l>$).

*Different inclusion probability at different frequency-locking ratios*

Each group exhibited frequency-locking ratios in differing degrees. Fig. 5 shows each group's inclusion probability at different frequency-locking ratios. If a subject had frequency-locking episodes with differing ratios, then the subject was included in each frequency-locking ratio. We used a two-proportion z-test to compare the difference of inclusion probabilities at a given frequency-locking ratio between two populations. Here, $H_0: P_x = P_y$ and $H_A: P_x \neq P_y$, where $P_x$ and $P_y$ are probabilities of including subjects from each population. We performed four different comparisons; MO vs. FO, MO vs. FY, FO vs. FY, and MO vs. healthy subjects (pooled FO and FY). The significance level was adjusted at $\alpha=0.0125$ for Bonferroni correction.

We detected no statistically significant differences at the 1:2 frequency-locking ratio between MO vs. FO, MO vs. FY, and FO vs. FY [two-proportion z-test, *p>0.0125*; Fig. 5A]. Although the inclusion probability of MO was almost double those of FO and FY, there was not enough statistical power to distinguish the difference between them because of the small sample sizes of FO and FY (at the 1:2 frequency-locking ratio). However, we detected significantly different inclusion probabilities between MO vs. healthy subjects (pooled FO and FY) at this ratio [two-proportion z-test, *p=5.53e-3*; Fig. 5A]. At the 1:3 frequency-locking ratio, we observed no differences in inclusion probabilities for all four comparisons [two-proportion z-test, *p>0.0125*; Fig. 5B]. At the 1:4 frequency-locking ratio, the differences in inclusion probabilities between MO and healthy subjects (pooled FO and FY) were very prominent (Fig. 5C). At this ratio, inclusion probabilities between MO vs. FO, MO vs. FY, and MO vs. healthy subjects (pooled FO and FY) were significantly different [two-proportion z-test, *p≤3.44e-4*; Fig. 5C]. Note that the inclusion probabilities of each FO and FY were almost thrice that of MO.

Thus, while healthy subjects tended to exhibit more of the 1:4 frequency-locking ratio, diseased subjects exhibited more of the 1:2 frequency-locking ratio. Though the fine temporal structure of synchronized dynamics exhibited statistically significant variations between groups, all cases preferred short desynchronizations.

**Discussion**

*Invariant and variable properties of cardiorespiratory synchronization*

In this study, we analyzed the fine temporal structure of the synchrony between cardiac and respiratory rhythms in healthy young subjects, healthy elderly subjects, and elderly subjects with coronary artery disease. As was observed earlier, cardiac and respiratory rhythms showed many different frequency-locking ratios, and these ratios changed over time (4, 16, 24, 31, 34, 35). We found significant differences between subject groups in the tendency to exhibit specific frequency-locking ratios and in the temporal patterns of synchronizations/desynchronizations. As we mentioned in the Introduction, certain temporal dynamical properties of cardiac dynamics remain intact regardless of physiological conditions (6, 36). Similarly, heart rate's influence on respiration does not differ with age or some physiological states (14). In this study, we discover other invariant properties within the cardiorespiratory system. The synchronized dynamics of the cardiorespiratory system have a very specific temporal pattern of synchrony: the oscillations go out of synchrony frequently, but return to the synchronous state quickly, regardless of age or disease status.

We observed significant differences in temporal patterns of synchronization between three subject groups (healthy young and elderly subjects, and elderly subjects with coronary artery disease), within each frequency-locking ratio. Healthy young subjects had stronger synchrony than those of healthy elderly subjects at the 1:3 and 1:4 frequency-locking ratios. These findings are consistent with previous results of reduced cardiorespiratory coupling with age in healthy subjects (4, 14). In contrast, at the 1:2 frequency-locking ratio, elderly subjects with coronary artery disease displayed higher levels of synchrony and shorter desynchronization episodes than healthy subjects of both young and old ages. Note that 1:5 and 1:6 frequency ratio dynamics were not synchronous and 1:1 dynamics was observed only in few subjects from the disease group, hence all three ratios (1:1, 1:5, and 1:6) were excluded from the analysis.

This exclusion prevents one from studying cardiorespiratory synchronization and RSA simultaneously. It was observed earlier that when frequency-locking ratios are high, RSA may be significant while cardiorespiratory phase-locking is weak (45). The strength of RSA depends on respiration frequency: it decreases as respiratory frequency increases (5, 18, 37, 39). At the frequency-locking ratios considered in our study (1:2, 1:3, 1:4), breathing rhythms were fast (>0.24 Hz) so that RSA is expected to be weak and not very significant. However, at higher ratios, when RSA is expected to be more prominent, cardiorespiratory synchronization is virtually absent (and thus the methods used here are not applicable).

It is also worth of noting that healthy subjects exhibited more of the higher (1:4) frequency-locking ratio, while elderly subjects with coronary artery disease exhibited more of the lower (1:2) frequency-locking ratio. Elderly subjects with coronary artery disease exhibited lower respiratory to heartbeat ratios and higher levels of cardiorespiratory synchrony. Thus, this group may be less adaptable to internal and external factors.

In spite of these age/disease-dependent effects, all groups clearly favored short desynchronization episodes (to differing degrees) regardless of age and disease status. The groups considered (especially MGH/MF) may be inhomogeneous, making it hard to draw conclusions about some specific patterns of cardiorespiratory synchrony. However, short desynchronization durations persisted in this subject group in a very robust way, and thus inhomogeneity of the studied group is unconcerning.

*Potential functional implications of observed synchronization properties*

Very strong synchrony implies a high probability of short desynchronization events because the oscillations spend less time in a desynchronous state. However, the oscillations examined in this study were only weakly synchronous. Thus, an important question is whether the cardiorespiratory system favors a state of weak synchrony with frequent short desynchronizations because this is the only option for *any* coupled oscillators. Ahn et al. (1) showed that coupled oscillators can exhibit a broad spectrum of desynchronization event durations with similar levels of synchrony. Therefore, the observed fine temporal structure of cardiorespiratory synchrony may have functional importance.

Here, we can speculate the potential functional importance of short desynchronization dynamics. The function of cardiorespiratory coordination is still debated. For example, it is not clear if RSA facilitates gas exchange (44). Coordination of the timing of heartbeats and phases of respiratory cycle (which is related to phase synchrony) was suggested to be essential to optimize pulmonary gas exchange, but recent studies suggest that the cardio-ventilatory coupling does not play a significant role in this optimization, at least in humans (38). Nevertheless, it is reasonable to suppose that phase synchronization of cardiac and respiratory rhythms is important for efficiency and coordination of some cardiac and respiratory functions. If this is truly so, the short desynchronization dynamics may further facilitate this coordination and its efficiency. Short desynchronization dynamics suggest that a synchronous state is easy to create, but also easy to break. These easy-to-create and easy-to-break properties are the feature of cardiorespiratory coordination at rest (as we ascertain in this manuscript).

As noted above, the complexity is reduced with pathological heart conditions (28). Relatively high synchrony levels, especially at the lower frequency ratios, may be pathological, as they reduce natural variability of cardiac and respiratory dynamics, which may be quite large. Low synchrony levels with prolonged synchronization and desynchronization intervals may act similarly due to the long length of synchronous episodes. Short desynchronization dynamics may be a way to facilitate the mutual coordination of cardiac and respiratory rhythms by constantly creating and breaking apart synchronous episodes. Thus, these short desynchronization dynamics may be an efficient mechanism by which the cardiorespiratory system adapts to internal changes and external perturbations.

The discovery of frequent short desynchronizations may fit well with the observations that disease and age promote stronger synchrony and lower frequency ratios. The latter two correspond to stronger coordination between two rhythms. Therefore, the conjectured role of short desynchronizations to facilitate dynamic changes in coordination may be quite important in disease and older age, where rhythms are more connected.

Interestingly, studies of EEG of healthy individuals (2) and studies of subcortical neural activity in Parkinsonian patients (27) revealed somewhat similar patterns of neural synchrony with persistent short desynchronization events. In these subjects, the probability of the shortest desynchronization and average duration of synchronization episodes depends on the brain area, the execution of a particular task, and disease status (2, 27), but overall they tend to be higher than the values observed here. Hence, neural synchrony exhibits more prominent short desynchronization dynamics than cardiorespiratory synchronization. A possible reason for this is that cortical and basal ganglia neurons may need to engage in quick formation and break-up of neural assemblies. Conversely, the cardiorespiratory system may benefit from slower, intermittent coordination of cardiac and respiratory rhythms.

**Acknowledgments**

We acknowledge Mathematical Biosciences Institute support and thank Dr. Yaroslav Molkov for his comments on the manuscript.

**Disclosures**

No conflicts of interest are declared by the authors.

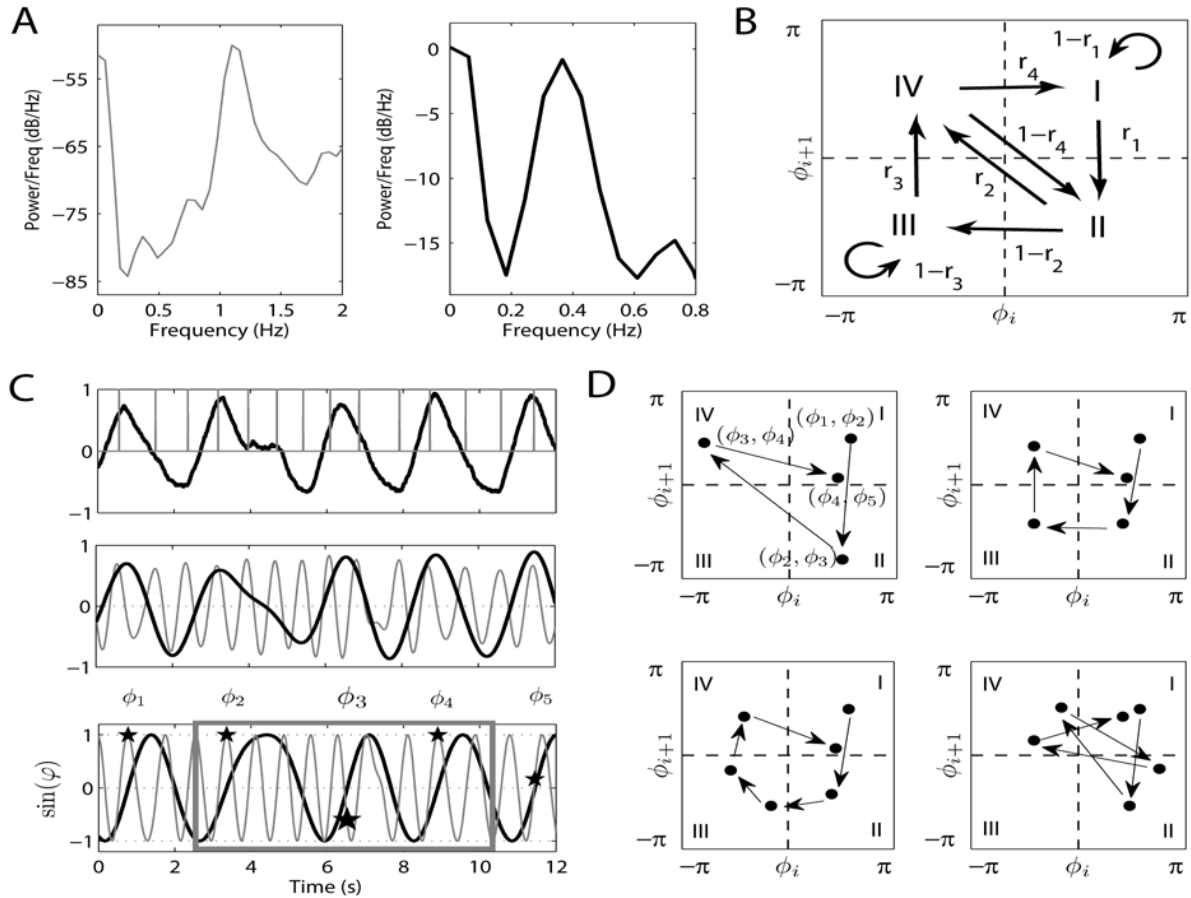

Figure 1. (A) Examples of PSDs from the ECG (left) and respiratory impedance (right) from a subject in the Fantasia young group. (B) Diagram of the $(\phi_i, \phi_{i+1})$ first-return map. The arrows indicate all possible transitions from one region to another and $r_{1,2,3,4}$ indicate the corresponding transition rates. (C) Thin gray lines are ECG signals and thick black lines are respiratory signals. The top panel shows the raw signals, the middle panel shows the signals filtered at 0.565~1.695 Hz for the ECG and 0.185~0.555 Hz for the respiratory impedance, and the bottom panel shows the sines of the phases from both filtered signals. The stars indicate the phases of the ECG when the phases of the respiratory impedance cross from negative to positive values. The gray box represents a desynchronization event with a duration of one cycle (in sync - out of sync - in sync). All units on the y-axes are dimensionless. The peaks of sines of phases in the bottom panel are shifted from the peaks of filtered signals in the middle panel because the filtered signals have sine and cosine components. Letters $\phi_{1,2,3,4,5}$ in the bottom panel of (C) correspond to the letters in the upper left plot in (D). (D) Example diagrams of desynchronization events of length one cycle (upper left), two cycles (upper right), and three cycles (bottom left and right).

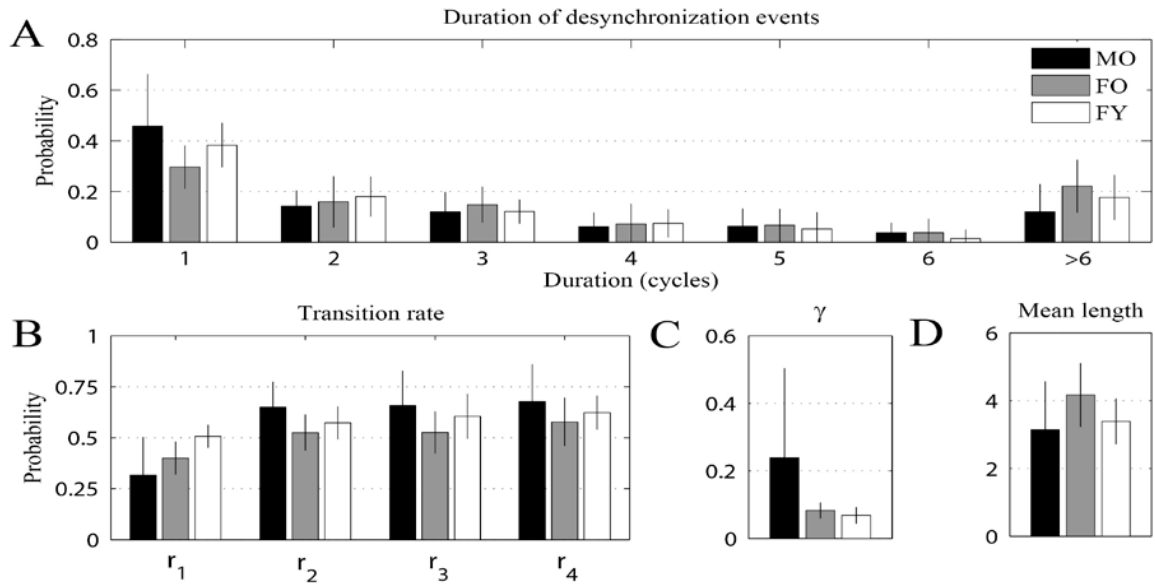

Figure 2. Dynamics of the cardiorespiratory synchronization at the 1:2 frequency-locking ratio for the three subject groups (MO, FO, and FY). (A) Distributions of desynchronization event durations. (B) Transition rates $r_{1,2,3,4}$. (C) Synchrony index $\gamma$. (D) Mean length of desynchronization event durations $<l>$. Mean $\pm$ SD is presented.

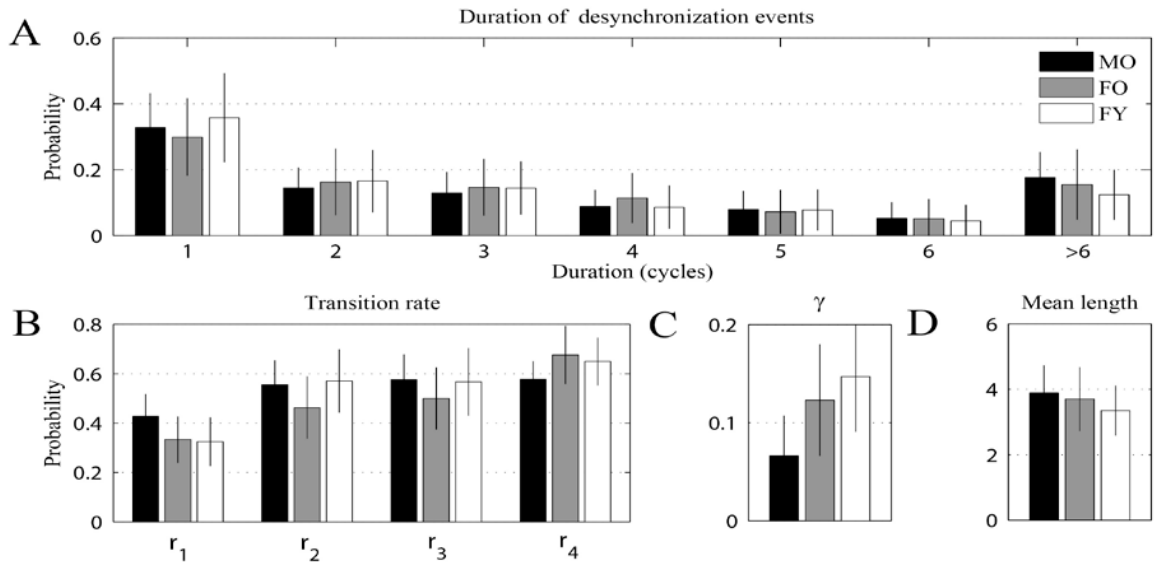

Figure 3. Dynamics of the cardiorespiratory synchronization at the 1:3 frequency-locking ratio for the three subject groups (MO, FO, and FY). (A) Distributions of desynchronization event durations. (B) Transition rates $r_{1,2,3,4}$. (C) Synchrony index $\gamma$. (D) Mean length of desynchronization event durations $<l>$. Mean $\pm$ SD is presented.

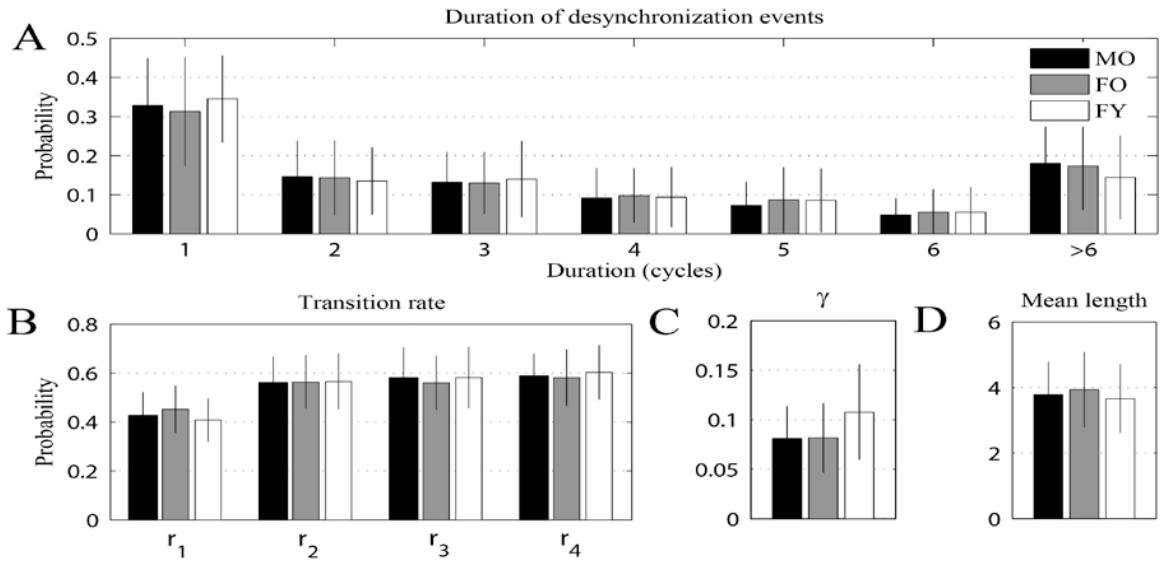

Figure 4. Dynamics of the cardiorespiratory synchronization at the 1:4 frequency-locking ratio for the three subject groups (MO, FO, and FY). (A) Distributions of desynchronization event durations. (B) Transition rates $r_{1,2,3,4}$. (C) Synchrony index $\gamma$. (D) Mean length of desynchronization event durations $<l>$. Mean $\pm$ SD is presented.

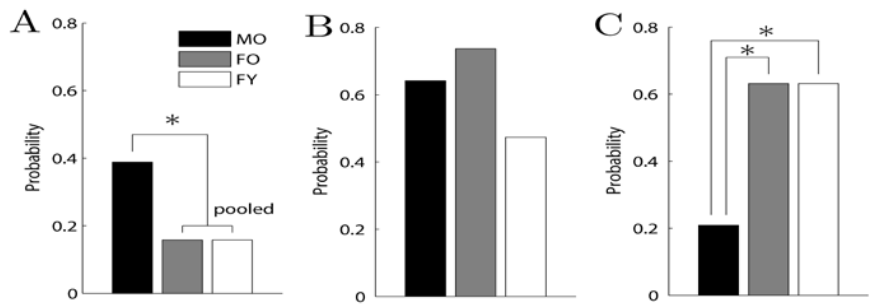

Figure 5. Inclusion probabilities of subjects for three groups at the frequency-locking ratios of (A) 1:2, (B) 1:3, and (C) 1:4. We compare the inclusion probabilities for MO vs. FO, MO vs. FY, FO vs. FY, and MO vs. healthy subjects (pooled FO and FY). ∗ Two-proportion z-test, *p<0.0125*.